\documentclass[twocolumn,aps,showpacs,preprintnumbers,lettersize]{revtex4}
\usepackage{amsfonts}
\usepackage{amsmath}
\usepackage{amssymb}
\usepackage{graphicx}

\begin{document}

\title{Mott Transition, Antiferromagnetism, and d-wave Superconductivity in
Two-Dimensional Organic Conductors}
\author{B. Kyung and A. -M. S. Tremblay}
\affiliation{D\'{e}partement de physique and Regroupement qu\'{e}b\'{e}cois sur les mat%
\'{e}riaux de pointe, Universit\'{e} de Sherbrooke, Sherbrooke, Qu\'{e}bec,
J1K 2R1, Canada}
\date{\today }

\begin{abstract}
We study the Mott transition, antiferromagnetism and superconductivity in
layered organic conductors using Cellular Dynamical Mean Field Theory for
the frustrated Hubbard model. A $d$-wave superconducting phase appears
between an antiferromagnetic insulator and a metal for $t^{\prime
}/t=0.3-0.7 $, or between a nonmagnetic Mott insulator (spin liquid) and a
metal for $t^{\prime }/t\geq 0.8$, in agreement with experiments on layered
organic conductors including $\kappa $-(ET)$_{2}$Cu$_{2}$(CN)$_{3}$. These
phases are separated by a strong first order transition. The phase diagram
gives much insight into the mechanism for d-wave superconductivity. Two
predictions are made.
\end{abstract}

\maketitle


Strong electronic correlations lead to fascinating phenomena such as
high-temperature superconductivity and metal-insulator (Mott) transitions.
In that context, layered organic conductors $\kappa $-(BEDT-TTF)$_{2}$X (X
denotes an anion) play a special role. They share many features with
high-temperature superconductors (HTSC) \cite{McKenzie:1997,Singleton:2002}, such as the
existence of d-wave superconductivity and antiferromagnetism. But they have
an even richer phase diagram since they also display a Mott transition and
possibly a $T=0$ spin liquid phase. Microscopically, they are described by
the two-dimensional one-band Hubbard model, just as HTSC, albeit on an
anisotropic triangular lattice instead of a square lattice.~\cite{Kuroki:2006}
The various phases of HTSC are explored essentially by changing doping. The phases of
organic conductors on the other hand are controlled by pressure and by
frustration, two variables over which one has very little control in the
HTSC. In this paper, we will see that we can capitalize on recent
theoretical progress on the theory of HTSC to understand organic conductors.
In return, the agreement that we find between theory and experiment for
these compounds leads to much insight into the origin of d-wave
superconductivity in the one-band Hubbard model in general.

Experimentally, metallic, paramagnetic insulating (spin liquid SL),
antiferromagnetic (AF), and unconventional superconducting (SC) phases are
all found, for example, in the pressure vs temperature phase diagram of
X=Cu[N(CN)$_{2}$]Cl~\cite{Lefebvre:2000}. The broken-symmetry states overlap
through a first order boundary that merges with the first order line of the
Mott metal-insulator transition in the normal state. Various experiments
suggest that the superconductivity has line nodes \cite{Arai:2001} that are
strongly suggestive of d-wave character. Changing the anion modifies the
frustration in the lattice. The anion (X=Cu$_{2}$(CN)$_{3}$)~\cite%
{Shimizu:2003}, which corresponds to large frustration, has attracted
considerable attention because of the transition from d-wave SC to a
possible SL state where no magnetic long-range order was found down to a
very low temperature ($23$mK), in contrast to other layered organic
materials.

The intriguing behaviors observed in layered organic conductors have
prompted theoretical studies using various analytical and numerical methods.
The existence of magnetic, metallic and spin liquid phases, but not d-wave
superconductivity, has been explored using a path-integral renormalization
group method for the frustrated Hubbard model by Morita \textit{et al.}~\cite%
{Morita:2002}. Signatures of a finite temperature Mott critical point in
agreement with experimental studies of $\kappa $-organics and with single
site DMFT have been found with Cellular Dynamical Mean Field Theory (CDMFT)~%
\cite{KSPB:2001} in conjunction with a quantum Monte Carlo (QMC) method, by
Parcollet \textit{et al.}~\cite{PBK:2004}.
At $T=0,$ using a variational Monte Carlo (VMC) method, Liu \textit{et al.}~%
\cite{Liu:2005} showed an unconventional SC ground state, sandwiched between
a conventional metal at weak coupling and a spin liquid at large coupling,
but antiferromagnetism was not considered. Similarly, with a U(1) gauge
theory in the slave-rotor representation, Lee \textit{et al.}~\cite{Lee:2005}
found a first order transition from a superconductor to a spin liquid and
Powell \textit{et al.}~\cite{Powell:2005} showed a first order transition
from a Mott insulator to a $d$-wave superconductor using a Gutzwiller
projection method for the Hubbard-Heisenberg model. Gan \textit{et al.}~\cite%
{Gan:2005} found a Gossamer superconductor at small $U$ and an AF insulator
at large $U$, separated by a first order transition. Only the variational
study of Watanabe \textit{et al.}~\cite{Watanabe:2006} considered the
possibility of all phases, metallic, d-wave SC, AF and SL. However, they did
not find AF where it is observed \cite{Lefebvre:2000} experimentally.
%

The layered organic conductors are described by the frustrated
two-dimensional Hubbard model~\cite{Kino:1996} 
\begin{equation}
H=\sum_{\langle ij\rangle ,\sigma }t_{ij}c_{i\sigma }^{\dagger }c_{j\sigma
}+U\sum_{i}n_{i\uparrow }n_{i\downarrow }-\mu \sum_{i\sigma }c_{i\sigma
}^{\dagger }c_{i\sigma },  \label{eq10}
\end{equation}%
where $c_{i\sigma }^{\dagger }$ ($c_{i\sigma }$) are creation (annihilation)
operators for electrons of spin $\sigma $, $n_{i\sigma }=c_{i\sigma
}^{\dagger }c_{i\sigma }$ is the density of $\sigma $ spin electrons. The
hopping amplitude $t_{ij}$ determines the anisotropic bare dispersion $%
\varepsilon _{\vec{k}}=-2t(\cos k_{x}+\cos k_{y})-2t^{\prime }\cos
(k_{x}+k_{y})$. (The HTSC have an additional $2t^{\prime }\cos (k_{x}-k_{y})$
in their dispersion). $U$ is the on-site repulsive interaction and $\mu $ is
the chemical potential controlling the electron density. Physically relevant
parameters for layered organic conductors are $U/t=5-10$ at half-filling ($%
n=1$) and $t^{\prime }/t=0.5-1.1$.

\begin{figure}[t]
\includegraphics[width=8.0cm]{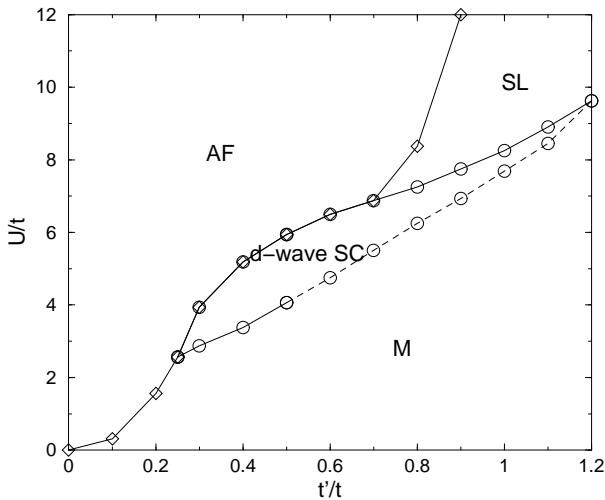}
\caption{Phase diagram of the frustrated two-dimensional Hubbard model at
zero temperature in the $U/t$ vs $t^{\prime }/t$ plane. AF, SL, SC and M
denote antiferromagnetic, spin liquid (nonmagnetic insulator),  
d-wave superconducting and metallic phases.}
\label{phase_diagram.fig}
\end{figure}

Our results are summarized by the $U/t$ vs $t^{\prime }/t$ phase diagram in
Fig.~\ref{phase_diagram.fig}. A $d$-wave SC phase appears in a relatively
narrow region between metallic (M) and AF phases, or between metallic (M)
and spin liquid phases (SL). 
The phase transition lines are all first order, except for the SC-metal
transition that is also first order but eventually turns second order
(dashed line) for sufficiently large $t^{\prime }/t$ (see caveat \cite{exotic_phase}). Three phases, M-AF-SC,
meet at the triple point near $t^{\prime }/t=0.25$ and $U/t=2.75,$
satisfying the Gibbs phase rule. There is a jump for all order parameters so
that this point does not exhibit SO(5) symmetry~\cite{Zhang:1997}. The SL
phase (paramagnetic Mott insulator) appears only for large $t^{\prime }/t$
(large frustration) and large $U/t$ (strong correlations). If we do not
allow for broken symmetry, we obtain a normal state that has a first order
Mott metal-insulator transition that coincides with the first order line
that separates SC from either AF or SL. That first order line however ends
at a critical point for $t^{\prime }/t$ close to $0.5.$

\textit{Predictions : }Based on the calculated phase diagram, we predict a
class of new materials (with $t^{\prime }/t\sim 0.8-0.9$) which would
undergo a series of phase transitions from an AF insulator to a paramagnetic
Mott insulator (spin liquid) to a $d$-wave superconductor to a metal with
increasing pressure. Also, we predict that if AF and SC are destroyed with
an external symmetry-restoring perturbation (magnetic field), the
first-order transition that separates them should remain as a first-order
Mott metal-insulator transition. 
In the following, we describe the method and results that lead to the above
phase diagram. We then discuss the remarkable agreement with existing
experiments. 


\textit{Method : }We use Cellular Dynamical Mean-Field Theory (CDMFT)~\cite%
{KSPB:2001}, a cluster approach that allows one to extend Dynamical Mean
Field Theory (DMFT) to include momentum dependence of the self-energy. 
CDMFT has been benchmarked and is accurate even in one dimension \cite%
{BKK:2003,CCKCK:2004}. The infinite lattice is tiled with identical clusters
of size $N_{c}$, and the degrees of freedom in the cluster are treated
exactly while the remaining ones are replaced by a bath of non-interacting
electrons that is determined self-consistently. To solve the quantum cluster
embedded in an effective SC medium, we consider a cluster-bath Hamiltonian
of the form~\cite{KKSTCK:2005,KCCKSKT:2005} 
\begin{eqnarray}
H &=&\sum_{\langle \mu \nu \rangle ,\sigma }t_{\mu \nu }c_{\mu \sigma
}^{\dagger }c_{\nu \sigma }+U\sum_{\mu }n_{\mu \uparrow }n_{\mu \downarrow }
\notag \\
+ &&\sum_{m,\sigma ,\alpha }\varepsilon _{m\sigma }^{\alpha }a_{m\sigma
}^{\dagger \alpha }a_{m\sigma }^{\alpha }+\sum_{m,\mu ,\sigma ,\alpha
}V_{m\mu \sigma }^{\alpha }(a_{m\sigma }^{\dagger \alpha }c_{\mu \sigma }+%
\mathrm{H.c.})  \notag \\
+ &&\sum_{m,n,\alpha }\Delta _{mn}(a_{m\uparrow }^{\alpha }a_{n\downarrow
}^{\alpha }+\mathrm{H.c.})\,.  \label{eq20}
\end{eqnarray}%
Here the indices $\mu ,\nu =1,\cdots ,N_{c}$ label sites within the cluster,
and $c_{\mu \sigma }$ and $a_{m\sigma }^{\alpha }$ annihilate electrons on
the cluster and the bath, respectively. $t_{\mu \nu }$ is the hopping matrix
within the cluster, and $\varepsilon _{m\sigma }^{\alpha }$ are the bath
energies and $V_{m\mu \sigma }^{\alpha }$ are the bath-cluster hybridization
matrices. $\Delta _{mn}$ represents the amplitude of SC correlations on the
bath with a given gap symmetry. Because superconductivity and
antiferromagnetism are allowed to compete on equal footing, $\varepsilon
_{m\sigma }^{\alpha }$ and $V_{m\mu \sigma }^{\alpha }$ carry a spin
variable $\sigma $ explicitly. In the present study we used $N_{c}=4$ sites
for the cluster and $N_{b}=8$ sites for the bath with $m=1,...,4$, $\alpha
=1,2$. In the normal state, because of symmetry, there are $8$ independent
parameters, while there are $18$ when AF and SC are allowed to compete. To
deal with superconductivity, the Nambu spinor representation is used for the
cluster operators so that the Weiss field, the cluster Green's function and
self-energy constructed from these operators are $8\times 8$ matrices. The
exact diagonalization method~\cite{CK:1994} is used to solve the
cluster-bath Hamiltonian Eq.~\ref{eq20} at zero temperature, which has the
advantage of computing dynamical quantities directly in real frequency and
of treating the large $U$ regime without difficulty. Although the present
study focuses on a $2\times 2$ cluster with additional $8$ bath sites, we
expect our results to be robust with respect to an increase in the cluster
size. This was verified by our recent low (but finite) temperature CDMFT+QMC
calculations~\cite{KKT:2005} where at intermediate to strong coupling a $%
2\times 2$ cluster accounts for more than $95\%$ of the correlation effect
of the infinite size cluster in the single-particle spectrum. Recent Variational Cluster Perturbation Theory (VCPT)
calculations~\cite{Sahebsara:2006} for the same Hamiltonian also confirmed
that results on a $2\times 2$ cluster are quantitatively similar to those on
larger clusters.


\begin{figure}[t]
\includegraphics[width=9.0cm]{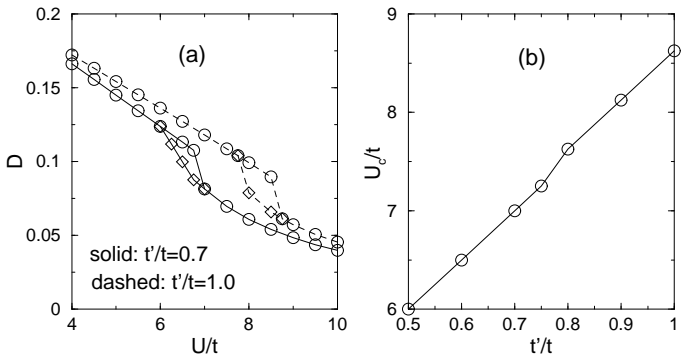}
\caption{(a) double occupancy $D=\langle n_{\uparrow}n_{\downarrow} \rangle$
as a function of $U/t$ for $t^{\prime}/t=0.7$ (solid curve) and $%
t^{\prime}/t=1.0$ (dashed curve), (b) critical $U_{c}/t$ as a function of $%
t^{\prime}/t$ in the normal state. The diamond symbols in (a) are obtained with the insulating
solution as an initial guess.}
\label{first_order.fig}
\end{figure}

\textit{Normal state Mott transition: }We first present the evidence of a
first order line of metal-to-insulator transition in the normal state. That
line also turns out to coincide with the phase boundary between AF and SC
phases for $t^{\prime }/t>0.5$. The present results obtained in the normal
state at $T=0$ would be relevant when the temperature is slightly higher than $T_{c}$
and $T_{N}$, or in regions where broken-symmetries are destroyed at zero
temperature by an external symmetry restoring perturbation. The first plot
in Fig.~\ref{first_order.fig}(a) shows the double occupancy $\langle
n_{\uparrow }n_{\downarrow }\rangle $ (circles) for two different $t^{\prime
}/t$. For $t^{\prime }/t=0.7$ it jumps discontinuously to a lower value near 
$U/t=7.0$, while for $t^{\prime }/t=1.0$ the first order transition occurs
near $U/t=8.5$ which is comparable to the value obtained by Parcollet 
\textit{et al.}~\cite{PBK:2004} in CDMFT+QMC and Watanabe \textit{et al.}~%
\cite{Watanabe:2006} in VMC. To show hysteresis associated with a first
order transition, we also calculated the double occupancy (diamonds in Fig.~%
\ref{first_order.fig}(a)) with the insulating solution as an initial guess. 
In this case the first order transition occurs for a value of $U$ that is
typically smaller by $0.5-0.75t$. A very similar hysteresis is found for all
the first order transitions shown in this paper, but for the rest of the
paper we show results obtained only with the metallic solution as an initial
guess. The critical $U_{c}$ shown in Fig.~\ref{first_order.fig}(b) decreases
monotonically with decreasing $t^{\prime }/t$ until the first order
transition disappears near $t^{\prime }/t=0.5$, close to $t^{\prime }/t\leq
0.4$ found by Watanabe \textit{et al.}~\cite{Watanabe:2006}.

\begin{figure}[t]
\includegraphics[width=5.0cm,angle=270]{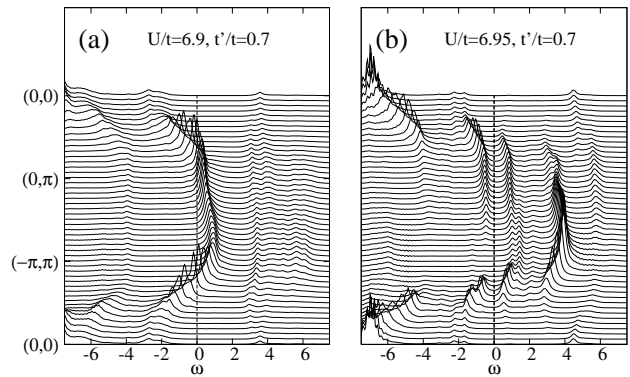}
\caption{Spectral function $A(\vec{k},\protect\omega)$ along some symmetry
directions for (a) $U/t=6.9$ and (b) $U/t=6.95$ for $t^{\prime}/t=0.7$
in the normal state.}
\label{A_k_w.fig}
\end{figure}


While various studies~\cite%
{Morita:2002,PBK:2004,Watanabe:2006,Lee:2005,Powell:2005,Gan:2005} suggested
the existence of a first order transition for the frustrated Hubbard model
in static quantities, here we present the strongest evidence of a first
order transition from a metal to an insulator by presenting dynamical
quantities $A(\vec{k},\omega )$. For $U/t=6.9$ slightly smaller than
critical $U_{c}/t$ for $t^{\prime }/t=0.7$, the spectral function has a
sharp quasiparticle peak near the Fermi wave vectors (Fig.~\ref{A_k_w.fig}%
(a)) consistent with a Fermi liquid picture. It is more evident from the
fact that $A(\vec{k},\omega )$ becomes sharper as $\vec{k}$ approaches the
Fermi surface. When $U/t$ is increased only by a tiny fraction (less than $%
1\%$), the first order metal-to-insulator transition manifests itself as the
massive reshuffling of the spectral weight. An insulating gap is present in
the whole Brillouin zone, reminiscent of the spectral function in the
two-dimensional Hubbard model at half-filling~\cite{KKSTCK:2005} where the
low energy bands inside the Hubbard bands are caused by short-range spin
correlations.

\textit{Broken symmetry states: }Next we study the existence of broken
symmetry states in the frustrated model and the role of the first order
transition in those states. 
\begin{figure}[t]
\includegraphics[width=9.0cm]{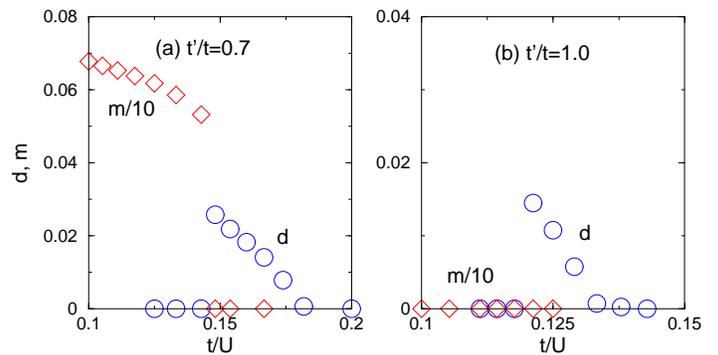}
\caption{(Color online) Order parameters of the frustrated two-dimensional Hubbard model as
a function of pressure (modeled as $t/U$) at zero temperature for (a) $%
t^{\prime }/t=0.7$ and (b) $t^{\prime }/t=1.0$. $d$ (circles) and $m$
(diamonds) denote the $d$-wave SC and AF order parameters, respectively. $m$
is multiplied by $0.1$ to fit in the plots.}
\label{phase_0.7+1.0.fig}
\end{figure}
Because of possible coexistence of AF and SC phases, the two order
parameters are treated on equal footing. For a SC phase two symmetries are
considered $d_{x^{2}-y^{2}}$ and $d_{xy}$. For all the parameters studied $%
d_{x^{2}-y^{2}}$ symmetry always dominates over $d_{xy}$ (see caveat \cite%
{gap_symmetry}). For $t^{\prime }/t=0.7$ (Fig.~\ref{phase_0.7+1.0.fig}(a))
AF and SC phases abruptly terminate near $U/t=7$, the same $U_{c}$ as in the
normal state where a metal-to-insulator transition occurs. 
We have not found evidence of homogeneous states with non-zero AF and SC
order parameters simultaneously, in contrast to the case of the square
lattice $t-t^{\prime }-U$ model for HTSC at finite doping, where AF can have
some itinerant character \cite{TKS:2005,Capone:2006}. 
With enough frustration, $t^{\prime }/t=1.0$ (Fig.~\ref{phase_0.7+1.0.fig}%
(b)) a direct first order transition from a paramagnetic Mott insulator to a
superconductor occurs near $U/t=8.25$ in agreement with other recent
theoretical works~\cite{Watanabe:2006,Lee:2005,Powell:2005}. That
paramagnetic Mott insulator is commonly refered to as a spin liquid. This
result is also in agreement with the study~\cite{Zheng:1999} of the $%
J-J^{\prime }$ Heisenberg model in which the N\'{e}el order persists up to $%
J^{\prime }/J=0.6-0.7$ ($t^{\prime }/t=0.77-0.84$). Nevertheless, we have to
leave open the possibility that magnetically ordered phases different from
AF are more stable than the SL.

\textit{Contact with experiment: }In first approximation, the effect of
pressure is to decrease the ratio $U/t,$ \cite{McKenzie:1997} even though $%
t^{\prime }/t$ should also change slightly \cite{Campos:1996}. We now
demonstrate that our phase diagram Fig.~\ref{phase_diagram.fig} accounts for
a surprisingly large number of experimental results if we assume that
pressure only decreases the $U/t$ ratio at fixed $t^{\prime }/t$, the
various compounds being associated to a given $t^{\prime }/t$. For $%
t^{\prime }/t=0.7$ (Fig.~\ref{phase_0.7+1.0.fig}(a)) which is relevant to $%
\kappa $-(ET)$_{2}$Cu[N(CN)$_{2}$]Cl, Fig.~\ref{phase_diagram.fig} shows
that, as observed experimentally \cite{Lefebvre:2000}: (a) as pressure
increases, $\left( U/t\text{ decreases}\right) $, one crosses a first order
transition between AF and SC phase (b) the maximum SC order parameter occurs
at the phase boundary (c) in the normal state there is a first order
Mott-insulator transition, at essentially the same $t/U$ ratio. The d-wave
SC in our phase diagram however exists in a relatively small region of $t/U$
in Fig.~\ref{phase_0.7+1.0.fig}. In the experimental phase diagram~\cite%
{Lefebvre:2000} the SC region extends far beyond the AF+SC phase boundary in
the pressure vs temperature plane, but in the absence of a precise scale
connecting pressure and $t/U$ this cannot be taken as a real disagreement.
For $t^{\prime }/t=1.0,$ close to the value for $\kappa $-(ET)$_{2}$Cu$_{2}$%
(CN)$_{3}$, the AF phase is not stabilized even for large $U/t$ due to too
strong frustration and the transition is between a SL and SC, (Fig.~\ref%
{phase_0.7+1.0.fig}(b)) consistent with recent experiments by Shimizu 
\textit{et al.}~\cite{Shimizu:2003}.

Another clear overall trend in our results is that the maximum value of the
SC order parameter decreases monotonically with $t^{\prime }/t$ (0.0276,
0.0258, 0.0228, 0.0186, 0.0145 for $t^{\prime }/t=0.6,0.7,0.8,0.9,1.0$).
This trend is remarkably consistent with experiments where compounds with
weaker frustration (smaller $t^{\prime }/t$) have higher $T_{c}$, for
instance, \cite{McKenzie:1997,Watanabe:2006} $T_{c}=11.6,10.4,3.9$K for for
X=Cu[N(CN)$_{2}$]Br($t^{\prime }/t=0.68$), X=Cu(NCS)$_{2}$($t^{\prime
}/t=0.84$), X=Cu$_{2}$(CN)$_{3}$($t^{\prime }/t=1.06$), respectively.

\textit{Mechanism for d-wave superconductivity: }
Along the AF-SC and SL-SC phase transition lines, the maximum value of the
SC order parameter increases with $U$ until $U\sim 6t$ and then it starts to
decrease, as does the value of $J=4t^{2}/U.$ In previous studies of HTSC,%
\textit{\ }\cite{KCCKSKT:2005,SLMT:2005} the maximum value of the doping
dependent SC order parameter was found to scale similarly with interaction
strength. In the organics, at fixed $t^{\prime }/t$ the SC order parameter
always increases with increasing $U$ until it drops to zero at the phase
boundary. The SC region is broadest near $t^{\prime }/t=0.5$ and becomes
narrower for both smaller and larger $t^{\prime }/t$, vanishing near $%
t^{\prime }/t=0.25$ and $t^{\prime }/t=1.2$, respectively. If there is not
enough frustration in the system, as in the case of near perfect nesting at
small $t^{\prime }/t,$ AF long-range order is stabilized, suppressing the SC
phase. Increasing frustration then helps d-wave SC, as suggested for the
cuprates \cite{Andersen95}. However, too strong frustration at large $%
t^{\prime }/t$ suppresses even short-range singlet correlations on which
pairing correlations may build up so that SC disappears again for $t^{\prime
}/t\geq 1.2$. Note that the ratio of interaction strength $U$ to hopping
integral $t$ is of the same order in the organics and in HTSC but that the
energy scale set by the hopping integrals is an order of magnitude smaller
in these compounds than in the HTSC.

In conclusion, we obtained the phase diagram for layered organic conductors
using CDMFT. The calculated sequence of phases and the nature of the
transitions between them are consistent with observations in that class of
compounds. In addition, the observed maximum value of $T_{c}$ near the Mott
transition is consistent with the calculated maximum SC order parameter.
This allows us to predict a new class of materials. In conjunction with
results for HTSC \cite{SLMT:2005,KCCKSKT:2005}, the present results also
allow us to understand more deeply the role of exchange interactions and of
frustration for d-wave superconductivity.

We thank C. Bourbonnais, M. Poirier, P. Sahebsara, D.~S\'{e}n\'{e}chal, 
and M. Tanatar
for useful discussions, and J. Schmalian for pointing out several recent
references on organic conductors. Computations were performed on the Elix2
Beowulf cluster and on the Dell cluster of the RQCHP. The present work was
supported by NSERC (Canada), FQRNT (Qu\'{e}bec), CFI (Canada), CIAR, the
Tier I Canada Research Chair Program (A.-M.S.T.).



\begin{thebibliography}{99}
\bibitem{exotic_phase} Where the transition changes from 1st to 2nd order
there might be some extra more-exotic phases that we are not
able to resolve on a 2X2 cluster.
\bibitem{gap_symmetry} The possibility of $d_{x^{2}-y^{2}} + id_{xy}$ gap
symmetry cannot be ruled out.
\end{thebibliography}
\end{document}